\documentclass[aps,amsmath,amssymb,twocolumn,10pt,pra]{revtex4-2}
\usepackage[T1]{fontenc}
\usepackage{graphicx}  
\usepackage[colorlinks=true,linkcolor=blue,citecolor=red,urlcolor=magenta]{hyperref}
\usepackage[dvipsnames]{xcolor}
\usepackage{charter}

\begin{document}  
  
\title{Several fermions strongly interacting with a heavy mobile impurity\\ in a one-dimensional harmonic trap}
\author{Damian W{\l}odzy\'nski} 
\affiliation{Institute of Physics, Polish Academy of Sciences, Aleja Lotnik\'ow 32/46, PL-02668 Warsaw, Poland} 
\date{\today} 

\begin{abstract} 
We propose a numerically exact method for a mixture with a single impurity immersed in several majority fermions, confined in a harmonic potential. We separate one of the degrees of freedom through an appropriately tailored canonical transformation and perform exact diagonalization on the simplified Hamiltonian. This method is especially effective for a heavy impurity, where it outmatches the typical exact diagonalization approach. We used our method to calculate energy and density profiles of the first few eigenstates for the mixture with up to ten majority fermions.
\end{abstract}  
\maketitle

\section{Introduction} \label{Sec1}

Recent progress in experiments with ultracold atoms has allowed for physical realizations of seemingly purely theoretical models. Notable examples are few-body fermionic mixtures produced in Heidelberg \cite{Tunable, Fermionization, FewToMany}. In these experiments, \textsuperscript{6}Li atoms, in two different hyperfine states, have been trapped in a quasi-one-dimensional harmonic potential. Atomic species interact by contact interactions, which can be tuned using Feshbach resonance. The number of atoms in the mixture has also been precisely controlled. These experiments are the motivation for the further theoretical studies of few-body ultracold mixtures in one dimension (see \cite{few_review} for a review).

In the present work, we analyze a special case of a fermionic mixture with a single particle immersed in $N$ majority fermions. In the many-body limit (an impurity in a Fermi sea), this is known as a  Fermi polaron problem \cite{Massignan_2014}. In the few-body case, this problem has been extensively studied for equal masses of components. For a mixture in uniform potential, the eigenproblem has been solved analytically, both for repulsive \cite{McGuire1} and attractive interactions \cite{McGuire2}. For the trapped systems, especially important are mixtures in harmonic confinement since they were obtained experimentally \cite{FewToMany}. In the simplest two-body case ($N=1$), there exist an analytical solution \cite{Busch1998}. For larger mixtures, the problem has been analyzed numerically \cite{sciadv.1500197,PhysRevA.92.023634,PhysRevA.91.013620,PhysRevA.88.021602,Loft_2016,Lindgren_2014}. Other studied confinements include double-well potential \cite{PhysRevA.96.063603}, optical lattice \cite{Duncan_2017}, and a combined harmonic and periodic potential \cite{PhysRevA.89.053621}. Moreover, the mixture has been analyzed for an arbitrary confining potential in the regime of strong interactions \cite{inf_G,PhysRevA.91.023620,Volosniev2014}. In the case of mass imbalance,the problem has been addressed for the uniform potential \cite{https://doi.org/10.48550/arxiv.2112.06627}, mainly for the light impurity \cite{PhysRevA.89.052706,PhysRevA.106}.

In our analysis, we focus on the strongly interacting mixture in the harmonic confinement with the heavy impurity \cite{Mistakidis_2019,PhysRevA.101.053619}. We use a numerical method specifically designed to handle this problem. We perform dedicated canonical transformation on the Hamiltonian. The resulting Hamiltonian is split into the center-of-mass part, which has an analytical solution, and the relative part described as a system of weakly interacting fermions. The latter is solved using the exact diagonalization method.

This paper is organized as follows. The analyzed model is described in Section \ref{Sec2}. In Section \ref{Sec3}, we introduce the coordinate transformation used to simplify the problem. In Section \ref{Sec4} we present the numerical method used to solve the problem and compare it to the typical approach. The results for the first few stationary states of the system are presented in Section \ref{Sec5}. Finally, Section \ref{Sec6} contains our conclusions.

\section{The system studied} \label{Sec2}
In this paper, we consider a system of $N$ spinless fermions with mass $m$, interacting with a single impurity through contact interactions. The mixture is trapped in a one-dimensional harmonic trap with the same frequency $\Omega$ for both components. The analysis is performed in natural units of the harmonic oscillator for the majority fermions. Therefore units of energy, length, and interaction strength are $\hbar\Omega$, $\sqrt{\frac{\hbar}{m\Omega}}$, and $\sqrt{\frac{m}{\hbar^3\Omega}}$, respectively. The Hamiltonian of the system has a form:
\begin{align} \label{hamiltonian}
\begin{aligned}
\hat{\cal H} &= -\frac{1}{2M}\frac{\partial^2}{\partial y^2} + \frac{M}{2} y^2 +\\  &+ {\int \!\! dx\ } \hat{\psi}^\dagger(x)\left[ -\frac{1}{2}\frac{\partial^2}{\partial x^2} + \frac{1}{2} x^2 + g \delta(x-y) \right] \hat{\psi}(x)
\end{aligned}
\end{align}
where $M$ is the mass of the impurity expressed as a ratio to the mass of the other component. The Hamiltonian is partially written in the second quantization. The majority fermions are described with fermionic field operators $\hat{\psi}^\dagger(x)$ and $\hat{\psi}(x)$, which obey an anti-commutation relation $\{\hat{\psi}^\dagger(x),\hat{\psi}(x')\}=\delta(x-x')$. Therefore their fermionic statistic is incorporated into the Hamiltonian. For the impurity, we use the first quantization because it simplifies the coordinates transformation presented in the next section. The Hamiltonian in this mixed formalism acts on a tensor product of a single-particle Hilbert space in the position representation and a Fock space in the occupation number representation. In the analysis, we focus on the case of strong but finite repulsive interactions ($g\geq 1$) and a heavy impurity ($M > 1$).

Calculating properties of this system is a numerical challenge due to the strong interactions. One can approach this problem using the exact diagonalization method \cite{Weisse2008}. It allows for high accuracy but requires large amounts of computational resources. Therefore it is limited to few-body systems. On the other hand, this approach has a wide range of applications - it can be used for any mass ratio $M$, any trapping potential, as well as for the systems with more than one impurity. In this paper, we propose an alternative solution, designed specifically for systems with a single heavy impurity in uniform or quadratic potential.

\section{The transformation} \label{Sec3}

In the proposed method, we start by changing the coordinate system. Instead of the particle positions, we use the center-of-mass coordinate and the relative distances between the fermions and the impurity:
\begin{align} \label{new_coordinates}
y \rightarrow \hat{\cal S}y\hat{\cal S}^{-1} &= \frac{M}{M+N} y + \frac{1}{M+N} \hat{x}\\
x \rightarrow \hat{\cal S}\hat{x}\hat{\cal S}^{-1} &= \hat{x} - N y
\end{align}
New coordinates are obtained through the canonical transformation governed by the operator:
\begin{align} \label{transformation_op}
\hat{\cal S}(y,\partial_y) = \exp\left( -i y \hat{p} \right) \exp\left( \frac{\hat{x} \partial_y  }{M+N} \right)
\end{align}
where $\hat{x} = \int dx \hat{\psi}^\dagger(x) x \hat{\psi}(x)$ and $\hat{p}=-i\int dx \hat{\psi}^\dagger(x)\partial_x \hat{\psi}(x)$ are single-particle position and momentum operators for the majority fermions, respectively.
By using this transformation, we separate the Hamiltonian (\ref{hamiltonian}) into two parts:
\begin{align} \label{H_trans}
\hat{\cal H} \rightarrow \hat{\cal S}^{-1}\hat{\cal H}\hat{\cal S} &= \hat{\cal H}_y + \hat{\cal H}_x \ \ ,
\end{align}
where:
\begin{align}
 \hat{\cal H}_y &= -\frac{1}{2(M+N)}\frac{\partial^2}{\partial y^2} + \frac{M+N}{2} y^2
\end{align}
describes a harmonic oscillator with eigenstates:
\begin{align} \label{eigvec_CM}
\Phi_n(y) \propto H_n\!\left(\sqrt{M+N}y\right) e^{-\frac{(M+N) y^2}{2}}
\end{align}
($H_n()$ is n-th Hermite polynomial) and eigenenergies $E_n = n+\frac{1}{2}$. The second term describes a system of $N$ interacting fermions:

\begin{align} \label{hamiltonian_x}
\begin{aligned}
\hat{\cal H}_x &= {\int \!\! dx\ } \hat{\psi}^\dagger(x)h(x) \hat{\psi}(x) + \\
&+ {\int \!\! dx\ }  {\int \!\! dx'\ } \hat{\psi}^\dagger(x) \hat{\psi}^\dagger(x') V(x,x') \hat{\psi}(x') \hat{\psi}(x)
\end{aligned}
\end{align}
with single-particle part:
\begin{align} \label{busch}
h(x&) = -\frac{1}{2}\frac{M+1}{M}\frac{\partial^2}{\partial x^2} + \frac{1}{2}\frac{M+N-1}{M+N}x^2 + g\delta(x)
\end{align}
and an effective interaction:
\begin{align} \label{U}
V(x&,x')= - \frac{1}{2M}\frac{\partial^2}{\partial x \partial x'}-\frac{1}{2(M+N)} x x' 
\end{align}

This transformation can be viewed as an extension of the so-called Lee-Low-Pines (LLP) transformation \cite{LLP1, LLP2}, which also can be applied to the Fermi polaron problem \cite{LLP_example}. The LLP transformation separates relative motion, but the applicability is limited to the uniform potential. Our transformation generalizes this idea to any potential with a separable center of mass.

\section{The exact diagonalization} \label{Sec4}

	After the transformation is performed, the problem is reduced to the system of $N$ fermions described by the Hamiltonian $\hat{\cal H}_x$. The effective interaction between these fermions depends both on their positions and momenta. Moreover, the interaction term is inversely proportional to the mass ratio $M$. Therefore, in the case of heavy impurity, it is a weakly interacting system. In order to solve this problem, we use the exact diagonalization method. In this approach, we want to represent the Hamiltonian (\ref{hamiltonian_x}) as a matrix in the Fock basis of the noninteracting system. We begin by decomposing the field operator in a single-particle basis:
\begin{align} \label{field_operator}
\hat{\psi}(x)= \sum_m \hat{a}_m \phi_m(x) \ \ ,
\end{align}
where $\hat{a}_m$ are annihilation operators, and $\phi_m(x)$ are eigenfunctions of the single-particle part of the hamiltonian $\hat{\cal H}_x$:
\begin{align}
h(x) \phi_m(x) = \epsilon_m \phi_m(x) \ \ .
\end{align}
Fortunately, this eigenproblem can be solved analytically \cite{Busch1998}. Using an effective mass $\mu= \sqrt{\frac{M(M+N-1)}{(M+1)(M+N)}}$, the eigenfunctions can be written as:
\begin{align}
\phi_m(x) \propto  
\begin{cases}
       H_m\!\left(\sqrt{\mu}x\right) e^{-\frac{\mu x^2}{2}} & \text{if $m$ is odd}\ ,\\
       U\!\left(\frac{1-2\epsilon_m}{4},\frac{1}{2},\mu x^2\right) e^{-\frac{\mu x^2}{2}} & \text{if $m$ is even}
\end{cases} 
\end{align}
where $U()$ is the confluent hypergeometric function of the second kind. The eigenvalues are equal to $\epsilon_m= \sqrt{\frac{(M+1)(M+N-1)}{M(M+N)}} \nu_m$, where $\nu_m=m+\frac{1}{2}$ for odd $m$, while for even $m$, they are solutions of the equation:
\begin{align}
\frac{\Gamma\left(\frac{3-2\nu_m}{4}\right)}{\Gamma\left(\frac{1-2\nu_m}{4}\right)} = - \sqrt{\frac{M(M+N)}{(M+1)(M+N-1)}} \frac{\sqrt{\mu}g}{2}
\end{align}

After applying decomposition (\ref{field_operator}), we rewrite the Hamiltonian (\ref{hamiltonian_x}) to the form
\begin{align}
\hat{\cal H}_x &= \sum_m \epsilon_m \hat{a}^\dagger_m \hat{a}_m + \sum_{i j k l} V_{ijkl} \hat{a}^\dagger_i \hat{a}^\dagger_j \hat{a}_k \hat{a}_l
\end{align}
where $V_{ijkl}={\int \!\! dx\ }  {\int \!\! dx'\ } \hat{\psi}^\dagger(x) \hat{\psi}^\dagger(x') V(x,x') \hat{\psi}(x') \hat{\psi}(x)$. In this matrix representation of the hamiltonian $\hat{\cal H}_x$, every element corresponds to the Slater determinant of $N$ single-particle states. The matrix has the infinite number of elements, but high-energy states are numerically irrelevant when we calculate the low-energy many-body eigenstates.  Therefore we cut the sum in (\ref{field_operator}) on some index $K$. It is clear that, for a big enough cutoff $K$, the problem can be solved with any desired accuracy. As a result, the hamiltonian $\hat{\cal H}_x$ is reduced to a finite matrix that can be diagonalized. The overall spectrum of hamiltonian $\hat{\cal H}$ is obtained by combining the eigenvalues of this matrix with the center-of-mass energies $E_n$.

\begin{figure}
\centering
\includegraphics[width=\linewidth]{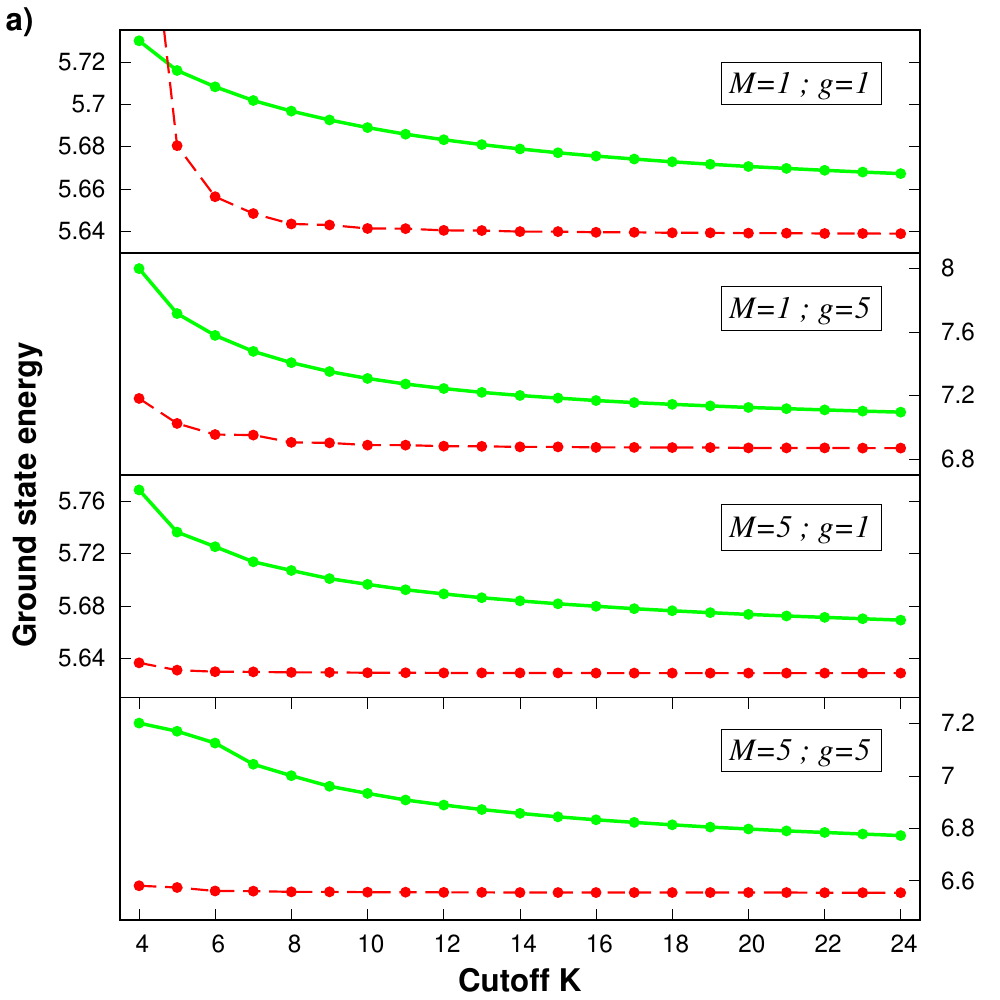} \\
\includegraphics[width=\linewidth]{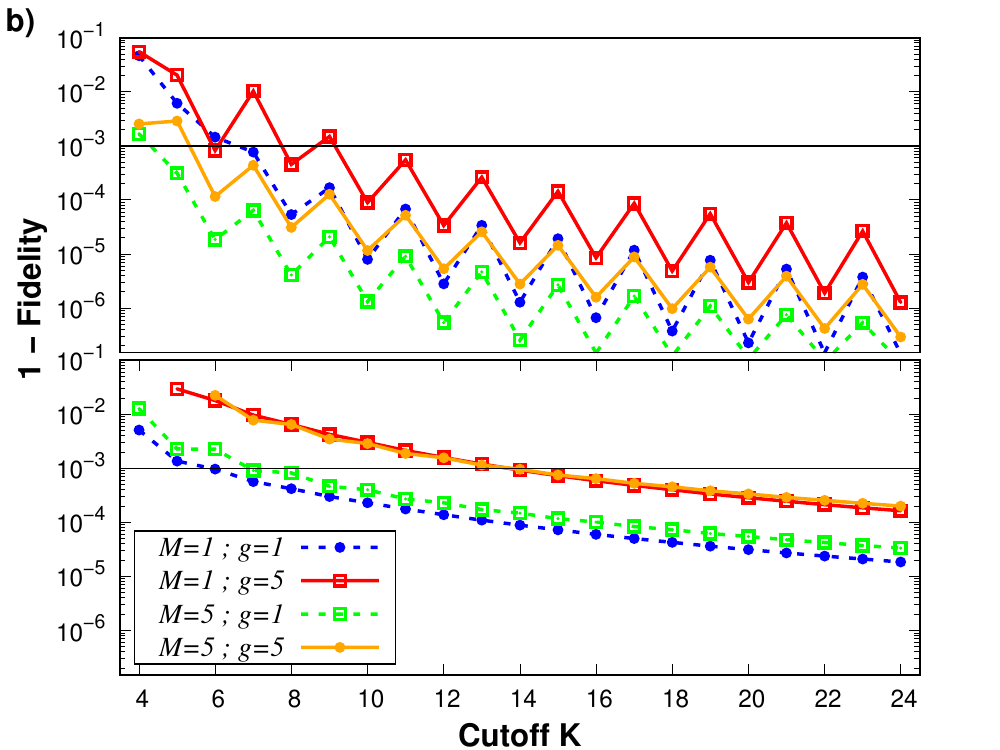}
\caption{ (a) Ground state energy in the $1+3$ system as a function of the state cutoff, for different values of mass ratio $M$ and interaction strength $g$. Results of exact diagonalization with and without the transformation are presented with dashed red and solid green lines, respectively. The energies are expressed in units of $\hbar\Omega$. (b) The fidelity for the same parameters. The top and bottom panels present the results for the new method and the default exact diagonalization, respectively.  \label{Fig1}}
\end{figure}

To evaluate how effective the new method is, we compare its predictions with the exact diagonalization used directly on the original Hamiltonian (\ref{hamiltonian}). In this standard approach, both components of the mixture are described in the second quantization with the field operators. The operators are decomposed in a single-particle basis (with the same cutoff $K$) and used to rewrite the Hamiltonian into a matrix which is then diagonalized. We check how accurately these two methods obtain the ground state of the system depending on the value of the cutoff. The results are presented in Fig \ref{Fig1}. The top figures show the ground state energy as a function of the cutoff $K$. The comparison is made for the system with $N=3$ fermions and four different sets of parameters $M$ and $g$. As expected, the results tend to decrease with increasing cutoff, as they approach the actual groundstate energy. It is also clear that, for all considered cases, the proposed method is much more accurate. That is the case both for heavy impurity and for the mass-balanced system. The reason is that accuracy of this method is less affected by the strength of the contact interactions since they are included directly in single-particle functions and not in the interaction term.  For the system with a very light impurity, the effective interaction term (\ref{U}) would be large and, therefore, the proposed method would be less effective.

For further comparison, we also calculate the fidelity:
\begin{align}
{\cal F}_K = \left|\langle \Psi^K | \Psi^{K+1} \rangle \right|
\end{align}
where $| \Psi^{K} \rangle$ is the eigenvector of the system with the cutoff $K$. This quantity describes how much the resulting state change when the cutoff increase. It is clear that as the eigenvector approaches the actual ground state, the fidelity should approach unity. For convenience, the results in Fig \ref{Fig1}b are presented as $1-{\cal F}_K$ such that they can be displayed in the logarithmic scale. The plot shows that the fidelity (as a function of the cutoff) approaches unity much faster in the proposed method. This again shows the advantage of this method. Moreover, the results become even better for larger impurity mass, which is not the case for the conventional exact diagonalization. On the other hand, stronger contact interactions decrease the accuracy of both methods. It should also be recalled that, in the proposed approach, one degree of freedom is calculated analytically. Therefore even for the same cutoff, the new method requires less computational resources than the typical exact diagonalization method.

\begin{figure}
\centering
\includegraphics[width=\linewidth]{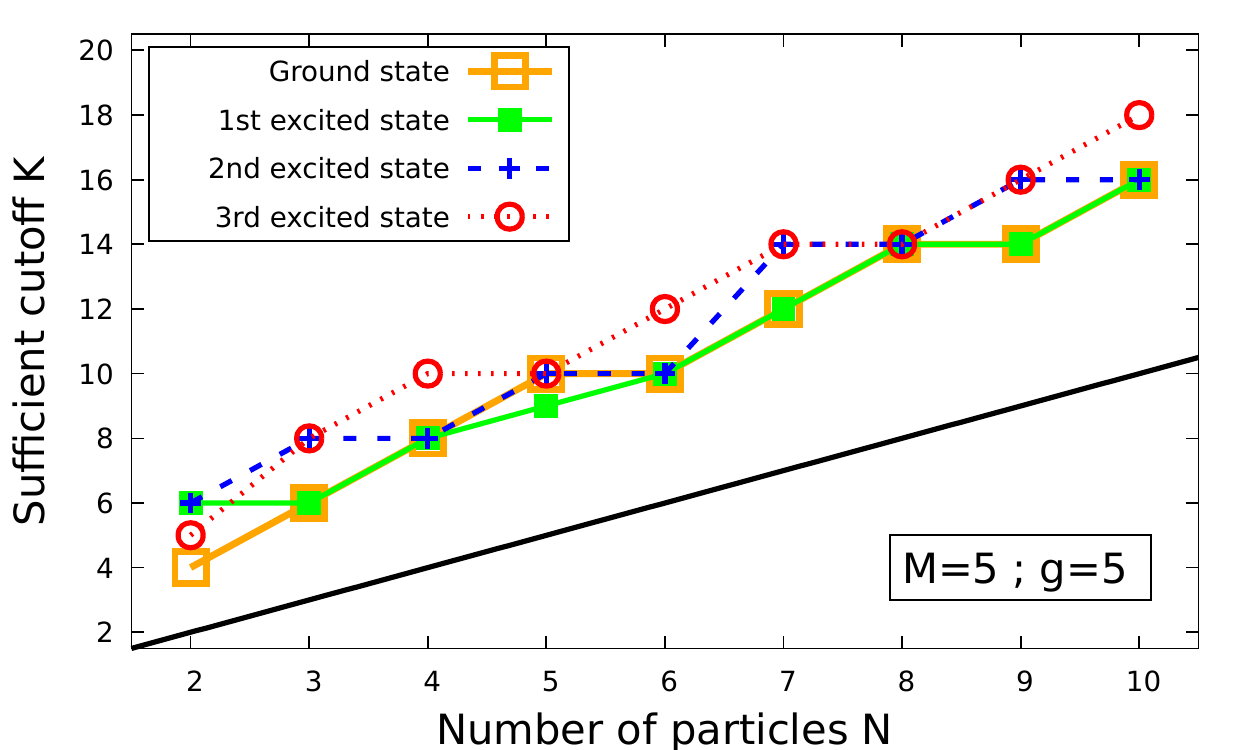}
\caption{ Sufficient cutoff (defined in the text) as a function of a particle's number. Four curves show results for the ground state and the first three excitations of the relative part of the hamiltonian (\ref{hamiltonian_x}). The black line corresponds to the lowest possible value for cutoffs ($K=N$). \label{Fig2}}
\end{figure}

Fidelity can be used to determine whether the cutoff is large enough for accurate numerical calculations. Here we assume that cutoff $K$ is sufficiently large when for every $k\geq K$, fidelity ${\cal F}_k$ is larger than $0.999$. Such sufficient cutoffs are plotted in Fig \ref{Fig2} for a system with heavy impurity and strong interactions ($M=5$ and $g=5$). Results were calculated for the ground state and the first few excitations of the relative motion. As one can see from the graph, both the ground state and the lowest excited state require a relatively low cutoff for accurate calculations (thanks to significant mass imbalance). It should be noted that choosing higher center-of-mass excitations does not influence the accuracy of the method, since this part of the Hamiltonian is always solved analytically. 

\begin{figure}
\centering
\includegraphics[width=\linewidth]{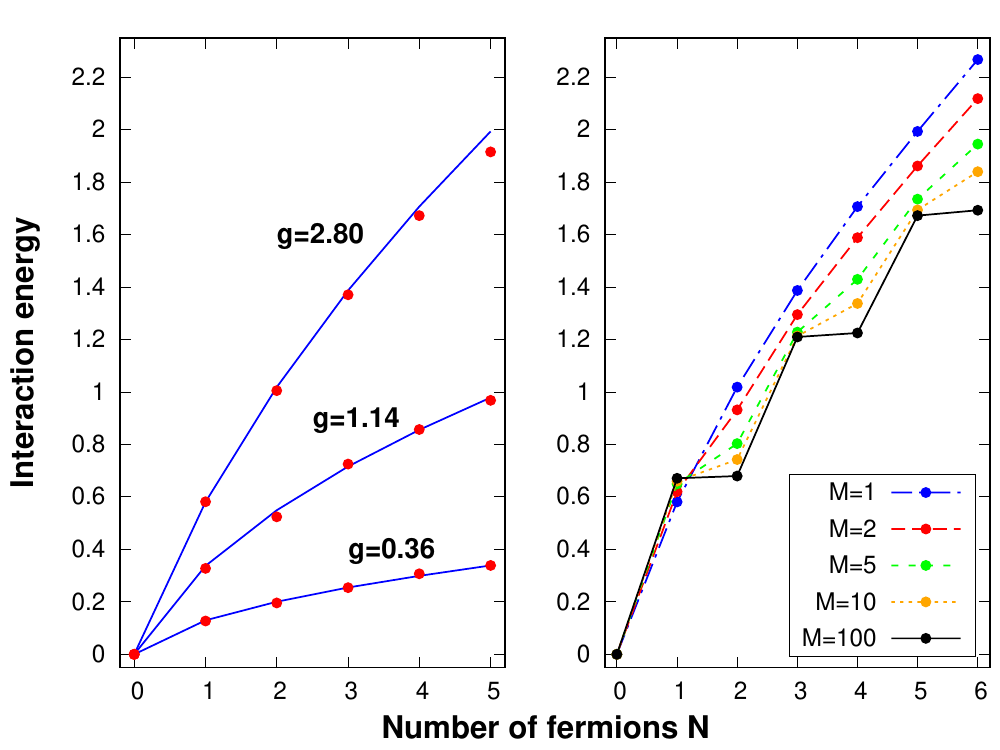}
\caption{ Interaction energy as a function of the particle's number. Left panel: different interaction strength in mass-balanced system ($M=1$). Experimantal data from \cite{FewToMany} (red dots) and numerical results (blue line). Right panel: different mass ratio $M$ for strong interactions ($g=2.80$). The energies are expressed in units of $\hbar\Omega$.\label{Fig3}}
\end{figure}

\section{Stationary properties} \label{Sec5}

Using our method, we calculate some stationary properties of the few-body fermi gas interacting with the impurity. We start with the ground state energy for the mass-balanced systems ($M=1$). In this case, there are available experimental data \cite{FewToMany}, which can be compared to the numerical results. In the experiment, the researchers measured the interaction energy instead of the ground state energy. Interaction energy is defined as a difference between the energy of the system with and without the inter-component interactions. The ground state energy of a non-interacting system is known analytically to be equal to $\frac{N^2+1}{2}$. Therefore the interaction energy is directly related to the ground state energy of the interacting system. A comparison of numerical results with experimental data is presented on the left panel of Fig \ref{Fig3}. Three lines correspond to weak ($g=0.36$), medium ($g=1.14$), and strong ($g=2.80$) interactions. As one can see from the graph, our results are in good agreement with the experimental data. It may seem that there is a significant relative error for $N=5$ and strong interactions. That is because we are comparing the interaction energy, which, in this case, is only a small part of the ground state energy that we calculate. For the ground state energy, the relative error is smaller than $1\%$ for all compared data.

On the right panel of Fig \ref{Fig3}, there is interaction energy calculated for the mass-imbalanced systems in the strong interaction regime ($g=2.80$). As can be seen from the graph, the interaction energy, as a function of the particle's number, changes qualitatively for a strong mass imbalance, from  a smooth curve to a step function. Increasing the number of fermions by one to the even values affects the interaction energy significantly less than with the odd values. This behavior can be understood using our transformation. In the limit of infinitely heavy impurity, the effective interaction term (\ref{U}) disappears. Therefore, the relative part of the hamiltonian reduces to:
\begin{align}
\hat{\cal H}_x \rightarrow {\int \!\! dx\ } \hat{\psi}^\dagger(x)\left(-\frac{1}{2}\frac{\partial^2}{\partial x^2} + \frac{1}{2}x^2 + g\delta(x) \right) \hat{\psi}(x) 
\end{align}
which describes non-interacting fermions in the harmonic trap with the additional delta potential in the center. In the ground state, the fermions occupy the first $N$ single-particle eigenstates. Every second such state is an odd function with respect to the spatial inverse $x\rightarrow-x$. Thus it is not affected by the Dirac potential in the middle. These single-particle states do not depend on the interaction strength $g$. Adding a new particle to the system with an even number of fermions increases groundstate energy in the same way regardless of whether there are interactions. Therefore interaction energy does not increase. In this limit, steps are ideally horizontal.

\begin{figure}
\centering
\includegraphics[width=\linewidth]{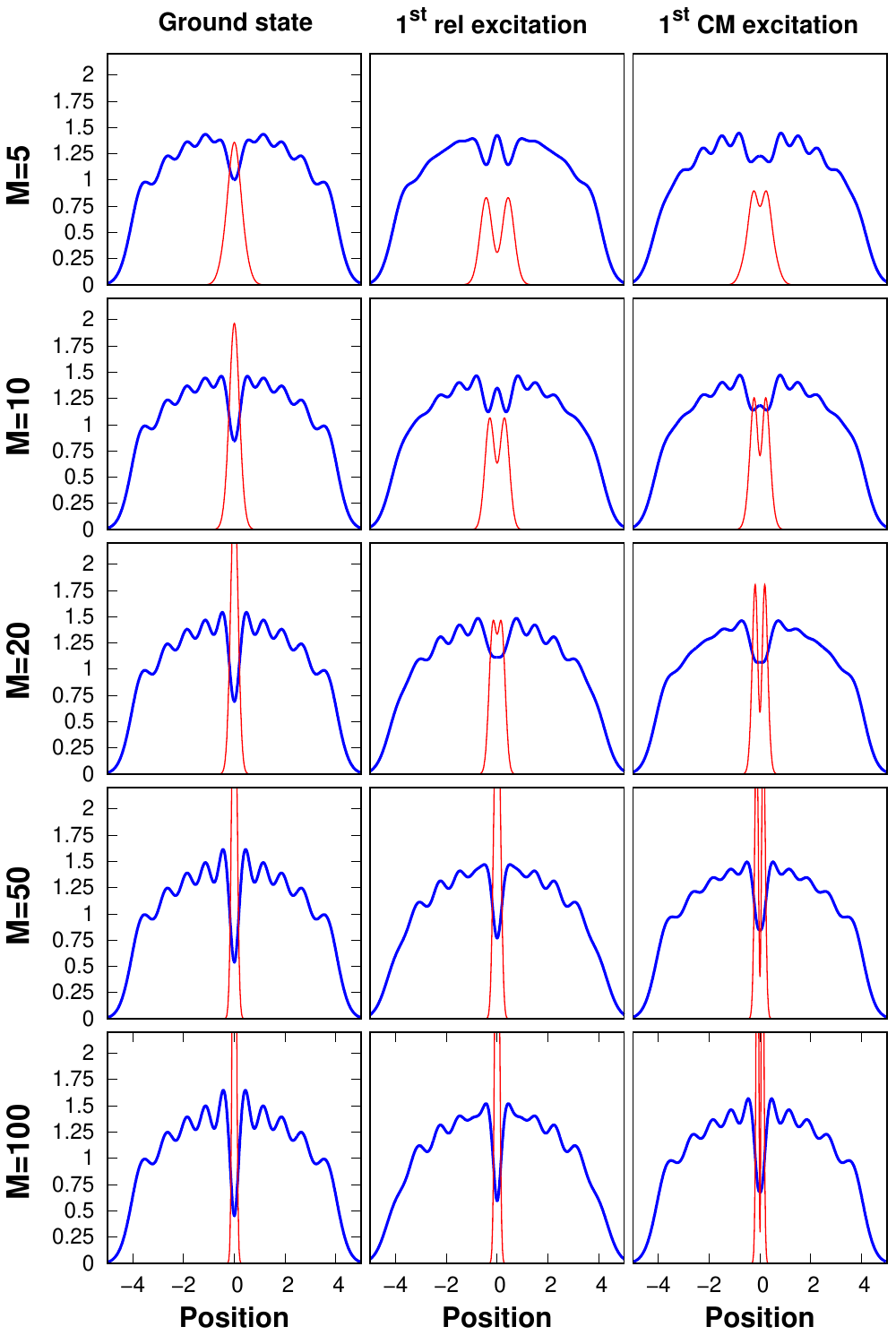}
\caption{Single-particle density profiles for the impurity and $N=10$ fermions (thin red and thick blue lines, respectively). 3 columns correspond to the ground state, the first excitation in relative motion and the first excitation in the center-of-mass motion. Results are presented for different mass ratios ($M=5,10,20,50$ and $100$). All for strong inter-component interactions ($g=5$). The densities and the positions are expressed in the units of $\sqrt{m\Omega/\hbar}$ and $\sqrt{\hbar/(m\Omega)}$, respectively. \label{Fig4}}
\end{figure}

Besides energies, one can also calculate density profiles. For the coordinates after the transformation, the density profiles of the eigenstate $| \Phi_n \rangle \otimes | \Omega_m \rangle$ (center-of-mass and relative part respectively) are equal:
\begin{align}
n_{\text {CM}}(y) &= |\Phi_n(y)|^2 \\ 
n_r(x) &= \Big|\Big| \hat{\psi}(x) | \Omega_m \rangle \Big|\Big|^2
\end{align}
where $\Phi_n(y)$ are functions from eq (\ref{eigvec_CM}).  Finding this quantity for the original Hamiltonian is not as simple as for the energy. The eigenstate has to be transformed back using the operator $\hat{\cal S}$:
\begin{align}
\label{transformed_back}
\begin{aligned}
\langle y| \hat{\cal S} | \Phi_n \rangle | \Omega_m \rangle &= e^{-iy\hat{p}} e^{\frac{\hat{x}\partial_y}{M+N}} \Phi_n(y) | \Omega_m \rangle = \\
&= e^{-iy\hat{p}}  \bar\Phi_n\!\!\left(y+\frac{\hat{x}}{M+N}\right) | \Omega_m \rangle
\end{aligned}
\end{align}
where $\bar\Phi_n\!()$ is an extension of the scalar function $\Phi_n\!()$ to the function acting on matrices. Since $\Phi_n\!()$ is an analytical function, this extension can be obtained rigorously by replacing scalars with matrices in the Taylor series. For the eigenstate in the original coordinates (\ref{transformed_back}), density profiles are equal:
\begin{align}
n_y(y) &= \bigg|\bigg|\bar\Phi_n\!\!\left(y+\frac{\hat{x}}{M+N}\right)\ | \Omega_m \rangle \bigg|\bigg|^2 \label{ny}\\ 
n_x(x) &= {\int \!\! dy\ } \bigg|\bigg| \hat{\psi}(x-y) \bar\Phi_n\!\!\left(y+\frac{\hat{x}}{M+N}\right)| \Omega_m \rangle \bigg|\bigg|^2 \label{nx}
\end{align}
where $\hat{\psi}(x-y)=e^{iy\hat{p}}\hat{\psi}(x) e^{-iy\hat{p}} $. This can be generalized further to many-body density profiles:
\begin{align}
\begin{aligned}
&n(x_1,...,x_k,y) = \\
&= \bigg|\bigg| \hat{\psi}(x_1\!-\!y)...\hat{\psi}(x_k\!-\!y) \bar\Phi_n\!\!\left(y\!+\!\frac{\hat{x}}{M\!+\!N}\right)| \Omega_m \rangle \bigg|\bigg|^2
\end{aligned}
\end{align}
From the numerical standpoint, one can calculate these quantities through diagonalization of the operator $\hat x$ in the Fock basis.

The single-particle density profiles for the system with $N=10$, interaction strength $g=5$ and for gradually increasing mass ratio $M$ are presented in Fig \ref{Fig4}. In the ground state (first column), the density profile for fermions doesn't change significantly as mass imbalance increases. Even for $M=5$, it is qualitatively the same as for $M=100$. Quantitatively, the most noticeable difference is that the local minimum at the center decreases as the mass imbalance increases. This is expected since the impurity becomes more and more localized in the middle of the trap and repulses the fermions from there. Surprisingly, the results for the excited states (second and third column) change qualitatively as mass imbalance increases. For $M=10$ there is still a local maximum in the middle of the trap, which is not the case in the heavy impurity limit. Even for $M=20$ shape of the density profile is significantly different than for $M=100$. Therefore finite-mass corrections are important even for the system with relatively heavy impurity, at least for the excited states.

\section{Conclusion} \label{Sec6}

In this paper, we've presented a new method to calculate the properties of the system of a few fermions interacting with a heavy impurity and confined in a harmonic trap. By comparing it to the standard approach of the exact diagonalization, we've shown that this new method allows for high accuracy of the results with smaller restrictions on the size of the system. Next, the method has been used to calculate basic stationary properties for the systems with up to $10$ fermions, namely the energies and the density profiles of the eigenstates.

It is worth noticing that this method can also be used to calculate the dynamic properties of the analyzed system. Once the initial state is transformed to the new coordinates, it can be decomposite into the Fock basis and evolve using any standard numerical method (for example Runge-Kutta algorithm). For the evolved state, the energy and density profiles can be calculated analogously to the eigenstates.

\bibliography{_BibTotal}

\begin{thebibliography}{29}%
\makeatletter
\providecommand \@ifxundefined [1]{%
 \@ifx{#1\undefined}
}%
\providecommand \@ifnum [1]{%
 \ifnum #1\expandafter \@firstoftwo
 \else \expandafter \@secondoftwo
 \fi
}%
\providecommand \@ifx [1]{%
 \ifx #1\expandafter \@firstoftwo
 \else \expandafter \@secondoftwo
 \fi
}%
\providecommand \natexlab [1]{#1}%
\providecommand \enquote  [1]{``#1''}%
\providecommand \bibnamefont  [1]{#1}%
\providecommand \bibfnamefont [1]{#1}%
\providecommand \citenamefont [1]{#1}%
\providecommand \href@noop [0]{\@secondoftwo}%
\providecommand \href [0]{\begingroup \@sanitize@url \@href}%
\providecommand \@href[1]{\@@startlink{#1}\@@href}%
\providecommand \@@href[1]{\endgroup#1\@@endlink}%
\providecommand \@sanitize@url [0]{\catcode `\\12\catcode `\$12\catcode
  `\&12\catcode `\#12\catcode `\^12\catcode `\_12\catcode `\%12\relax}%
\providecommand \@@startlink[1]{}%
\providecommand \@@endlink[0]{}%
\providecommand \url  [0]{\begingroup\@sanitize@url \@url }%
\providecommand \@url [1]{\endgroup\@href {#1}{\urlprefix }}%
\providecommand \urlprefix  [0]{URL }%
\providecommand \Eprint [0]{\href }%
\providecommand \doibase [0]{http://dx.doi.org/}%
\providecommand \selectlanguage [0]{\@gobble}%
\providecommand \bibinfo  [0]{\@secondoftwo}%
\providecommand \bibfield  [0]{\@secondoftwo}%
\providecommand \translation [1]{[#1]}%
\providecommand \BibitemOpen [0]{}%
\providecommand \bibitemStop [0]{}%
\providecommand \bibitemNoStop [0]{.\EOS\space}%
\providecommand \EOS [0]{\spacefactor3000\relax}%
\providecommand \BibitemShut  [1]{\csname bibitem#1\endcsname}%
\let\auto@bib@innerbib\@empty
\bibitem [{\citenamefont {Serwane}\ \emph {et~al.}(2011)\citenamefont
  {Serwane}, \citenamefont {Zürn}, \citenamefont {Lompe}, \citenamefont
  {Ottenstein}, \citenamefont {Wenz},\ and\ \citenamefont {Jochim}}]{Tunable}%
  \BibitemOpen
  \bibfield  {author} {\bibinfo {author} {\bibfnamefont {F.}~\bibnamefont
  {Serwane}}, \bibinfo {author} {\bibfnamefont {G.}~\bibnamefont {Zürn}},
  \bibinfo {author} {\bibfnamefont {T.}~\bibnamefont {Lompe}}, \bibinfo
  {author} {\bibfnamefont {T.~B.}\ \bibnamefont {Ottenstein}}, \bibinfo
  {author} {\bibfnamefont {A.~N.}\ \bibnamefont {Wenz}}, \ and\ \bibinfo
  {author} {\bibfnamefont {S.}~\bibnamefont {Jochim}},\ }\href {\doibase
  10.1126/science.1201351} {\bibfield  {journal} {\bibinfo  {journal}
  {Science}\ }\textbf {\bibinfo {volume} {332}},\ \bibinfo {pages} {336}
  (\bibinfo {year} {2011})}\BibitemShut {NoStop}%
\bibitem [{\citenamefont {Z\"urn}\ \emph {et~al.}(2012)\citenamefont {Z\"urn},
  \citenamefont {Serwane}, \citenamefont {Lompe}, \citenamefont {Wenz},
  \citenamefont {Ries}, \citenamefont {Bohn},\ and\ \citenamefont
  {Jochim}}]{Fermionization}%
  \BibitemOpen
  \bibfield  {author} {\bibinfo {author} {\bibfnamefont {G.}~\bibnamefont
  {Z\"urn}}, \bibinfo {author} {\bibfnamefont {F.}~\bibnamefont {Serwane}},
  \bibinfo {author} {\bibfnamefont {T.}~\bibnamefont {Lompe}}, \bibinfo
  {author} {\bibfnamefont {A.~N.}\ \bibnamefont {Wenz}}, \bibinfo {author}
  {\bibfnamefont {M.~G.}\ \bibnamefont {Ries}}, \bibinfo {author}
  {\bibfnamefont {J.~E.}\ \bibnamefont {Bohn}}, \ and\ \bibinfo {author}
  {\bibfnamefont {S.}~\bibnamefont {Jochim}},\ }\href {\doibase
  10.1103/PhysRevLett.108.075303} {\bibfield  {journal} {\bibinfo  {journal}
  {Phys. Rev. Lett.}\ }\textbf {\bibinfo {volume} {108}},\ \bibinfo {pages}
  {075303} (\bibinfo {year} {2012})}\BibitemShut {NoStop}%
\bibitem [{\citenamefont {Wenz}\ \emph {et~al.}(2013)\citenamefont {Wenz},
  \citenamefont {Zürn}, \citenamefont {Murmann}, \citenamefont {Brouzos},
  \citenamefont {Lompe},\ and\ \citenamefont {Jochim}}]{FewToMany}%
  \BibitemOpen
  \bibfield  {author} {\bibinfo {author} {\bibfnamefont {A.~N.}\ \bibnamefont
  {Wenz}}, \bibinfo {author} {\bibfnamefont {G.}~\bibnamefont {Zürn}},
  \bibinfo {author} {\bibfnamefont {S.}~\bibnamefont {Murmann}}, \bibinfo
  {author} {\bibfnamefont {I.}~\bibnamefont {Brouzos}}, \bibinfo {author}
  {\bibfnamefont {T.}~\bibnamefont {Lompe}}, \ and\ \bibinfo {author}
  {\bibfnamefont {S.}~\bibnamefont {Jochim}},\ }\href {\doibase
  10.1126/science.1240516} {\bibfield  {journal} {\bibinfo  {journal}
  {Science}\ }\textbf {\bibinfo {volume} {342}},\ \bibinfo {pages} {457}
  (\bibinfo {year} {2013})}\BibitemShut {NoStop}%
\bibitem [{\citenamefont {Sowi{\'{n}}ski}\ and\ \citenamefont
  {Garc{\'{\i}}a-March}(2019)}]{few_review}%
  \BibitemOpen
  \bibfield  {author} {\bibinfo {author} {\bibfnamefont {T.}~\bibnamefont
  {Sowi{\'{n}}ski}}\ and\ \bibinfo {author} {\bibfnamefont {M.~{\'{A}}.}\
  \bibnamefont {Garc{\'{\i}}a-March}},\ }\href {\doibase
  10.1088/1361-6633/ab3a80} {\bibfield  {journal} {\bibinfo  {journal} {Reports
  on Progress in Physics}\ }\textbf {\bibinfo {volume} {82}},\ \bibinfo {pages}
  {104401} (\bibinfo {year} {2019})}\BibitemShut {NoStop}%
\bibitem [{\citenamefont {Massignan}\ \emph {et~al.}(2014)\citenamefont
  {Massignan}, \citenamefont {Zaccanti},\ and\ \citenamefont
  {Bruun}}]{Massignan_2014}%
  \BibitemOpen
  \bibfield  {author} {\bibinfo {author} {\bibfnamefont {P.}~\bibnamefont
  {Massignan}}, \bibinfo {author} {\bibfnamefont {M.}~\bibnamefont {Zaccanti}},
  \ and\ \bibinfo {author} {\bibfnamefont {G.~M.}\ \bibnamefont {Bruun}},\
  }\href {\doibase 10.1088/0034-4885/77/3/034401} {\bibfield  {journal}
  {\bibinfo  {journal} {Reports on Progress in Physics}\ }\textbf {\bibinfo
  {volume} {77}},\ \bibinfo {pages} {034401} (\bibinfo {year}
  {2014})}\BibitemShut {NoStop}%
\bibitem [{\citenamefont {McGuire}(1965)}]{McGuire1}%
  \BibitemOpen
  \bibfield  {author} {\bibinfo {author} {\bibfnamefont {J.~B.}\ \bibnamefont
  {McGuire}},\ }\href {\doibase 10.1063/1.1704291} {\bibfield  {journal}
  {\bibinfo  {journal} {Journal of Mathematical Physics}\ }\textbf {\bibinfo
  {volume} {6}},\ \bibinfo {pages} {432} (\bibinfo {year} {1965})}\BibitemShut
  {NoStop}%
\bibitem [{\citenamefont {McGuire}(1966)}]{McGuire2}%
  \BibitemOpen
  \bibfield  {author} {\bibinfo {author} {\bibfnamefont {J.~B.}\ \bibnamefont
  {McGuire}},\ }\href {\doibase 10.1063/1.1704798} {\bibfield  {journal}
  {\bibinfo  {journal} {Journal of Mathematical Physics}\ }\textbf {\bibinfo
  {volume} {7}},\ \bibinfo {pages} {123} (\bibinfo {year} {1966})}\BibitemShut
  {NoStop}%
\bibitem [{\citenamefont {Busch}\ \emph {et~al.}(1998)\citenamefont {Busch},
  \citenamefont {Englert}, \citenamefont {Rza{\. z}ewski},\ and\ \citenamefont
  {Wilkens}}]{Busch1998}%
  \BibitemOpen
  \bibfield  {author} {\bibinfo {author} {\bibfnamefont {T.}~\bibnamefont
  {Busch}}, \bibinfo {author} {\bibfnamefont {B.-G.}\ \bibnamefont {Englert}},
  \bibinfo {author} {\bibfnamefont {K.}~\bibnamefont {Rza{\. z}ewski}}, \ and\
  \bibinfo {author} {\bibfnamefont {M.}~\bibnamefont {Wilkens}},\ }\href
  {\doibase 10.1023/A:1018705520999} {\bibfield  {journal} {\bibinfo  {journal}
  {Foundations of Physics}\ }\textbf {\bibinfo {volume} {28}},\ \bibinfo
  {pages} {549} (\bibinfo {year} {1998})}\BibitemShut {NoStop}%
\bibitem [{\citenamefont {Levinsen}\ \emph {et~al.}(2015)\citenamefont
  {Levinsen}, \citenamefont {Massignan}, \citenamefont {Bruun},\ and\
  \citenamefont {Parish}}]{sciadv.1500197}%
  \BibitemOpen
  \bibfield  {author} {\bibinfo {author} {\bibfnamefont {J.}~\bibnamefont
  {Levinsen}}, \bibinfo {author} {\bibfnamefont {P.}~\bibnamefont {Massignan}},
  \bibinfo {author} {\bibfnamefont {G.~M.}\ \bibnamefont {Bruun}}, \ and\
  \bibinfo {author} {\bibfnamefont {M.~M.}\ \bibnamefont {Parish}},\ }\href
  {\doibase 10.1126/sciadv.1500197} {\bibfield  {journal} {\bibinfo  {journal}
  {Science Advances}\ }\textbf {\bibinfo {volume} {1}},\ \bibinfo {pages}
  {e1500197} (\bibinfo {year} {2015})},\ \Eprint
  {http://arxiv.org/abs/https://www.science.org/doi/pdf/10.1126/sciadv.1500197}
  {https://www.science.org/doi/pdf/10.1126/sciadv.1500197} \BibitemShut
  {NoStop}%
\bibitem [{\citenamefont {Gra\ss{}}(2015)}]{PhysRevA.92.023634}%
  \BibitemOpen
  \bibfield  {author} {\bibinfo {author} {\bibfnamefont {T.}~\bibnamefont
  {Gra\ss{}}},\ }\href {\doibase 10.1103/PhysRevA.92.023634} {\bibfield
  {journal} {\bibinfo  {journal} {Phys. Rev. A}\ }\textbf {\bibinfo {volume}
  {92}},\ \bibinfo {pages} {023634} (\bibinfo {year} {2015})}\BibitemShut
  {NoStop}%
\bibitem [{\citenamefont {Gharashi}\ \emph {et~al.}(2015)\citenamefont
  {Gharashi}, \citenamefont {Yin}, \citenamefont {Yan},\ and\ \citenamefont
  {Blume}}]{PhysRevA.91.013620}%
  \BibitemOpen
  \bibfield  {author} {\bibinfo {author} {\bibfnamefont {S.~E.}\ \bibnamefont
  {Gharashi}}, \bibinfo {author} {\bibfnamefont {X.~Y.}\ \bibnamefont {Yin}},
  \bibinfo {author} {\bibfnamefont {Y.}~\bibnamefont {Yan}}, \ and\ \bibinfo
  {author} {\bibfnamefont {D.}~\bibnamefont {Blume}},\ }\href {\doibase
  10.1103/PhysRevA.91.013620} {\bibfield  {journal} {\bibinfo  {journal} {Phys.
  Rev. A}\ }\textbf {\bibinfo {volume} {91}},\ \bibinfo {pages} {013620}
  (\bibinfo {year} {2015})}\BibitemShut {NoStop}%
\bibitem [{\citenamefont {Astrakharchik}\ and\ \citenamefont
  {Brouzos}(2013)}]{PhysRevA.88.021602}%
  \BibitemOpen
  \bibfield  {author} {\bibinfo {author} {\bibfnamefont {G.~E.}\ \bibnamefont
  {Astrakharchik}}\ and\ \bibinfo {author} {\bibfnamefont {I.}~\bibnamefont
  {Brouzos}},\ }\href {\doibase 10.1103/PhysRevA.88.021602} {\bibfield
  {journal} {\bibinfo  {journal} {Phys. Rev. A}\ }\textbf {\bibinfo {volume}
  {88}},\ \bibinfo {pages} {021602} (\bibinfo {year} {2013})}\BibitemShut
  {NoStop}%
\bibitem [{\citenamefont {Loft}\ \emph {et~al.}(2016)\citenamefont {Loft},
  \citenamefont {Kristensen}, \citenamefont {Thomsen},\ and\ \citenamefont
  {Zinner}}]{Loft_2016}%
  \BibitemOpen
  \bibfield  {author} {\bibinfo {author} {\bibfnamefont {N.~J.~S.}\
  \bibnamefont {Loft}}, \bibinfo {author} {\bibfnamefont {L.~B.}\ \bibnamefont
  {Kristensen}}, \bibinfo {author} {\bibfnamefont {A.~E.}\ \bibnamefont
  {Thomsen}}, \ and\ \bibinfo {author} {\bibfnamefont {N.~T.}\ \bibnamefont
  {Zinner}},\ }\href {\doibase 10.1088/0953-4075/49/12/125305} {\bibfield
  {journal} {\bibinfo  {journal} {Journal of Physics B: Atomic, Molecular and
  Optical Physics}\ }\textbf {\bibinfo {volume} {49}},\ \bibinfo {pages}
  {125305} (\bibinfo {year} {2016})}\BibitemShut {NoStop}%
\bibitem [{\citenamefont {Lindgren}\ \emph {et~al.}(2014)\citenamefont
  {Lindgren}, \citenamefont {Rotureau}, \citenamefont {Forss{\'{e}}n},
  \citenamefont {Volosniev},\ and\ \citenamefont {Zinner}}]{Lindgren_2014}%
  \BibitemOpen
  \bibfield  {author} {\bibinfo {author} {\bibfnamefont {E.~J.}\ \bibnamefont
  {Lindgren}}, \bibinfo {author} {\bibfnamefont {J.}~\bibnamefont {Rotureau}},
  \bibinfo {author} {\bibfnamefont {C.}~\bibnamefont {Forss{\'{e}}n}}, \bibinfo
  {author} {\bibfnamefont {A.~G.}\ \bibnamefont {Volosniev}}, \ and\ \bibinfo
  {author} {\bibfnamefont {N.~T.}\ \bibnamefont {Zinner}},\ }\href {\doibase
  10.1088/1367-2630/16/6/063003} {\bibfield  {journal} {\bibinfo  {journal}
  {New Journal of Physics}\ }\textbf {\bibinfo {volume} {16}},\ \bibinfo
  {pages} {063003} (\bibinfo {year} {2014})}\BibitemShut {NoStop}%
\bibitem [{\citenamefont {Tylutki}\ \emph {et~al.}(2017)\citenamefont
  {Tylutki}, \citenamefont {Astrakharchik},\ and\ \citenamefont
  {Recati}}]{PhysRevA.96.063603}%
  \BibitemOpen
  \bibfield  {author} {\bibinfo {author} {\bibfnamefont {M.}~\bibnamefont
  {Tylutki}}, \bibinfo {author} {\bibfnamefont {G.~E.}\ \bibnamefont
  {Astrakharchik}}, \ and\ \bibinfo {author} {\bibfnamefont {A.}~\bibnamefont
  {Recati}},\ }\href {\doibase 10.1103/PhysRevA.96.063603} {\bibfield
  {journal} {\bibinfo  {journal} {Phys. Rev. A}\ }\textbf {\bibinfo {volume}
  {96}},\ \bibinfo {pages} {063603} (\bibinfo {year} {2017})}\BibitemShut
  {NoStop}%
\bibitem [{\citenamefont {Duncan}\ \emph {et~al.}(2017)\citenamefont {Duncan},
  \citenamefont {Bellotti}, \citenamefont {Öhberg}, \citenamefont {Zinner},\
  and\ \citenamefont {Valiente}}]{Duncan_2017}%
  \BibitemOpen
  \bibfield  {author} {\bibinfo {author} {\bibfnamefont {C.~W.}\ \bibnamefont
  {Duncan}}, \bibinfo {author} {\bibfnamefont {F.~F.}\ \bibnamefont
  {Bellotti}}, \bibinfo {author} {\bibfnamefont {P.}~\bibnamefont {Öhberg}},
  \bibinfo {author} {\bibfnamefont {N.~T.}\ \bibnamefont {Zinner}}, \ and\
  \bibinfo {author} {\bibfnamefont {M.}~\bibnamefont {Valiente}},\ }\href
  {\doibase 10.1088/1367-2630/aa753e} {\bibfield  {journal} {\bibinfo
  {journal} {New Journal of Physics}\ }\textbf {\bibinfo {volume} {19}},\
  \bibinfo {pages} {075001} (\bibinfo {year} {2017})}\BibitemShut {NoStop}%
\bibitem [{\citenamefont {Doggen}\ \emph {et~al.}(2014)\citenamefont {Doggen},
  \citenamefont {Korolyuk}, \citenamefont {T\"orm\"a},\ and\ \citenamefont
  {Kinnunen}}]{PhysRevA.89.053621}%
  \BibitemOpen
  \bibfield  {author} {\bibinfo {author} {\bibfnamefont {E.~V.~H.}\
  \bibnamefont {Doggen}}, \bibinfo {author} {\bibfnamefont {A.}~\bibnamefont
  {Korolyuk}}, \bibinfo {author} {\bibfnamefont {P.}~\bibnamefont {T\"orm\"a}},
  \ and\ \bibinfo {author} {\bibfnamefont {J.~J.}\ \bibnamefont {Kinnunen}},\
  }\href {\doibase 10.1103/PhysRevA.89.053621} {\bibfield  {journal} {\bibinfo
  {journal} {Phys. Rev. A}\ }\textbf {\bibinfo {volume} {89}},\ \bibinfo
  {pages} {053621} (\bibinfo {year} {2014})}\BibitemShut {NoStop}%
\bibitem [{\citenamefont {Guan}\ \emph {et~al.}(2009)\citenamefont {Guan},
  \citenamefont {Chen}, \citenamefont {Wang},\ and\ \citenamefont
  {Ma}}]{inf_G}%
  \BibitemOpen
  \bibfield  {author} {\bibinfo {author} {\bibfnamefont {L.}~\bibnamefont
  {Guan}}, \bibinfo {author} {\bibfnamefont {S.}~\bibnamefont {Chen}}, \bibinfo
  {author} {\bibfnamefont {Y.}~\bibnamefont {Wang}}, \ and\ \bibinfo {author}
  {\bibfnamefont {Z.-Q.}\ \bibnamefont {Ma}},\ }\href {\doibase
  10.1103/PhysRevLett.102.160402} {\bibfield  {journal} {\bibinfo  {journal}
  {Phys. Rev. Lett.}\ }\textbf {\bibinfo {volume} {102}},\ \bibinfo {pages}
  {160402} (\bibinfo {year} {2009})}\BibitemShut {NoStop}%
\bibitem [{\citenamefont {Volosniev}\ \emph {et~al.}(2015)\citenamefont
  {Volosniev}, \citenamefont {Petrosyan}, \citenamefont {Valiente},
  \citenamefont {Fedorov}, \citenamefont {Jensen},\ and\ \citenamefont
  {Zinner}}]{PhysRevA.91.023620}%
  \BibitemOpen
  \bibfield  {author} {\bibinfo {author} {\bibfnamefont {A.~G.}\ \bibnamefont
  {Volosniev}}, \bibinfo {author} {\bibfnamefont {D.}~\bibnamefont
  {Petrosyan}}, \bibinfo {author} {\bibfnamefont {M.}~\bibnamefont {Valiente}},
  \bibinfo {author} {\bibfnamefont {D.~V.}\ \bibnamefont {Fedorov}}, \bibinfo
  {author} {\bibfnamefont {A.~S.}\ \bibnamefont {Jensen}}, \ and\ \bibinfo
  {author} {\bibfnamefont {N.~T.}\ \bibnamefont {Zinner}},\ }\href {\doibase
  10.1103/PhysRevA.91.023620} {\bibfield  {journal} {\bibinfo  {journal} {Phys.
  Rev. A}\ }\textbf {\bibinfo {volume} {91}},\ \bibinfo {pages} {023620}
  (\bibinfo {year} {2015})}\BibitemShut {NoStop}%
\bibitem [{\citenamefont {Volosniev}\ \emph {et~al.}(2014)\citenamefont
  {Volosniev}, \citenamefont {Fedorov}, \citenamefont {Jensen}, \citenamefont
  {Valiente},\ and\ \citenamefont {Zinner}}]{Volosniev2014}%
  \BibitemOpen
  \bibfield  {author} {\bibinfo {author} {\bibfnamefont {A.~G.}\ \bibnamefont
  {Volosniev}}, \bibinfo {author} {\bibfnamefont {D.~V.}\ \bibnamefont
  {Fedorov}}, \bibinfo {author} {\bibfnamefont {A.~S.}\ \bibnamefont {Jensen}},
  \bibinfo {author} {\bibfnamefont {M.}~\bibnamefont {Valiente}}, \ and\
  \bibinfo {author} {\bibfnamefont {N.~T.}\ \bibnamefont {Zinner}},\ }\href
  {\doibase 10.1038/ncomms6300} {\bibfield  {journal} {\bibinfo  {journal}
  {Nature Communications}\ }\textbf {\bibinfo {volume} {5}},\ \bibinfo {pages}
  {5300} (\bibinfo {year} {2014})}\BibitemShut {NoStop}%
\bibitem [{\citenamefont {Burovski}\ \emph {et~al.}(2021)\citenamefont
  {Burovski}, \citenamefont {Gamayun},\ and\ \citenamefont
  {Lychkovskiy}}]{https://doi.org/10.48550/arxiv.2112.06627}%
  \BibitemOpen
  \bibfield  {author} {\bibinfo {author} {\bibfnamefont {E.}~\bibnamefont
  {Burovski}}, \bibinfo {author} {\bibfnamefont {O.}~\bibnamefont {Gamayun}}, \
  and\ \bibinfo {author} {\bibfnamefont {O.}~\bibnamefont {Lychkovskiy}},\
  }\href {\doibase 10.48550/ARXIV.2112.06627} {\enquote {\bibinfo {title}
  {Mobile impurity in a one-dimensional quantum gas: Exact diagonalization in
  the bethe ansatz basis},}\ } (\bibinfo {year} {2021}),\ \Eprint
  {http://arxiv.org/abs/arXiv:2112.06627} {arXiv:2112.06627} \BibitemShut
  {NoStop}%
\bibitem [{\citenamefont {Mehta}(2014)}]{PhysRevA.89.052706}%
  \BibitemOpen
  \bibfield  {author} {\bibinfo {author} {\bibfnamefont {N.~P.}\ \bibnamefont
  {Mehta}},\ }\href {\doibase 10.1103/PhysRevA.89.052706} {\bibfield  {journal}
  {\bibinfo  {journal} {Phys. Rev. A}\ }\textbf {\bibinfo {volume} {89}},\
  \bibinfo {pages} {052706} (\bibinfo {year} {2014})}\BibitemShut {NoStop}%
\bibitem [{\citenamefont {Tononi}\ \emph {et~al.}(2022)\citenamefont {Tononi},
  \citenamefont {Givois},\ and\ \citenamefont {Petrov}}]{PhysRevA.106}%
  \BibitemOpen
  \bibfield  {author} {\bibinfo {author} {\bibfnamefont {A.}~\bibnamefont
  {Tononi}}, \bibinfo {author} {\bibfnamefont {J.}~\bibnamefont {Givois}}, \
  and\ \bibinfo {author} {\bibfnamefont {D.~S.}\ \bibnamefont {Petrov}},\
  }\href {\doibase 10.1103/PhysRevA.106.L011302} {\bibfield  {journal}
  {\bibinfo  {journal} {Phys. Rev. A}\ }\textbf {\bibinfo {volume} {106}},\
  \bibinfo {pages} {L011302} (\bibinfo {year} {2022})}\BibitemShut {NoStop}%
\bibitem [{\citenamefont {Mistakidis}\ \emph {et~al.}(2019)\citenamefont
  {Mistakidis}, \citenamefont {Katsimiga}, \citenamefont {Koutentakis},\ and\
  \citenamefont {Schmelcher}}]{Mistakidis_2019}%
  \BibitemOpen
  \bibfield  {author} {\bibinfo {author} {\bibfnamefont {S.~I.}\ \bibnamefont
  {Mistakidis}}, \bibinfo {author} {\bibfnamefont {G.~C.}\ \bibnamefont
  {Katsimiga}}, \bibinfo {author} {\bibfnamefont {G.~M.}\ \bibnamefont
  {Koutentakis}}, \ and\ \bibinfo {author} {\bibfnamefont {P.}~\bibnamefont
  {Schmelcher}},\ }\href {\doibase 10.1088/1367-2630/ab1045} {\bibfield
  {journal} {\bibinfo  {journal} {New Journal of Physics}\ }\textbf {\bibinfo
  {volume} {21}},\ \bibinfo {pages} {043032} (\bibinfo {year}
  {2019})}\BibitemShut {NoStop}%
\bibitem [{\citenamefont {Kwasniok}\ \emph {et~al.}(2020)\citenamefont
  {Kwasniok}, \citenamefont {Mistakidis},\ and\ \citenamefont
  {Schmelcher}}]{PhysRevA.101.053619}%
  \BibitemOpen
  \bibfield  {author} {\bibinfo {author} {\bibfnamefont {J.}~\bibnamefont
  {Kwasniok}}, \bibinfo {author} {\bibfnamefont {S.~I.}\ \bibnamefont
  {Mistakidis}}, \ and\ \bibinfo {author} {\bibfnamefont {P.}~\bibnamefont
  {Schmelcher}},\ }\href {\doibase 10.1103/PhysRevA.101.053619} {\bibfield
  {journal} {\bibinfo  {journal} {Phys. Rev. A}\ }\textbf {\bibinfo {volume}
  {101}},\ \bibinfo {pages} {053619} (\bibinfo {year} {2020})}\BibitemShut
  {NoStop}%
\bibitem [{\citenamefont {Wei{\ss}e}\ and\ \citenamefont
  {Fehske}(2008)}]{Weisse2008}%
  \BibitemOpen
  \bibfield  {author} {\bibinfo {author} {\bibfnamefont {A.}~\bibnamefont
  {Wei{\ss}e}}\ and\ \bibinfo {author} {\bibfnamefont {H.}~\bibnamefont
  {Fehske}},\ }\enquote {\bibinfo {title} {Exact diagonalization techniques},}\
  in\ \href {\doibase 10.1007/978-3-540-74686-7_18} {\emph {\bibinfo
  {booktitle} {Computational Many-Particle Physics}}},\ \bibinfo {editor}
  {edited by\ \bibinfo {editor} {\bibfnamefont {H.}~\bibnamefont {Fehske}},
  \bibinfo {editor} {\bibfnamefont {R.}~\bibnamefont {Schneider}}, \ and\
  \bibinfo {editor} {\bibfnamefont {A.}~\bibnamefont {Wei{\ss}e}}}\ (\bibinfo
  {publisher} {Springer Berlin Heidelberg},\ \bibinfo {address} {Berlin,
  Heidelberg},\ \bibinfo {year} {2008})\ pp.\ \bibinfo {pages}
  {529--544}\BibitemShut {NoStop}%
\bibitem [{\citenamefont {Lee}\ \emph {et~al.}(1953)\citenamefont {Lee},
  \citenamefont {Low},\ and\ \citenamefont {Pines}}]{LLP1}%
  \BibitemOpen
  \bibfield  {author} {\bibinfo {author} {\bibfnamefont {T.~D.}\ \bibnamefont
  {Lee}}, \bibinfo {author} {\bibfnamefont {F.~E.}\ \bibnamefont {Low}}, \ and\
  \bibinfo {author} {\bibfnamefont {D.}~\bibnamefont {Pines}},\ }\href
  {\doibase 10.1103/PhysRev.90.297} {\bibfield  {journal} {\bibinfo  {journal}
  {Phys. Rev.}\ }\textbf {\bibinfo {volume} {90}},\ \bibinfo {pages} {297}
  (\bibinfo {year} {1953})}\BibitemShut {NoStop}%
\bibitem [{\citenamefont {Girardeau}(1983)}]{LLP2}%
  \BibitemOpen
  \bibfield  {author} {\bibinfo {author} {\bibfnamefont {M.~D.}\ \bibnamefont
  {Girardeau}},\ }\href {\doibase 10.1103/PhysRevA.28.3635} {\bibfield
  {journal} {\bibinfo  {journal} {Phys. Rev. A}\ }\textbf {\bibinfo {volume}
  {28}},\ \bibinfo {pages} {3635} (\bibinfo {year} {1983})}\BibitemShut
  {NoStop}%
\bibitem [{\citenamefont {Kain}\ and\ \citenamefont
  {Ling}(2017)}]{LLP_example}%
  \BibitemOpen
  \bibfield  {author} {\bibinfo {author} {\bibfnamefont {B.}~\bibnamefont
  {Kain}}\ and\ \bibinfo {author} {\bibfnamefont {H.~Y.}\ \bibnamefont
  {Ling}},\ }\href {\doibase 10.1103/PhysRevA.96.033627} {\bibfield  {journal}
  {\bibinfo  {journal} {Phys. Rev. A}\ }\textbf {\bibinfo {volume} {96}},\
  \bibinfo {pages} {033627} (\bibinfo {year} {2017})}\BibitemShut {NoStop}%
\end{thebibliography}%

\end{document}